\documentclass[11pt,twocolumn]{article}
\title{SIZE DEPENDENT CRUSH ANALYSIS OF LITHIUM ORTHOSILICATE PEBBLES\footnote{Published in Fusion Science and Technology, vol. 66, pages:136--141 (2014), doi:10.13182/FST13-737}}
\author{R.~K.~Annabattula,$^{a,\dagger}$ M.~Kolb,$^{b}$ Y.~Gan,$^{c}$ R.~Rolli$^{b}$ and M.~Kamlah$^{b}$\\
	$^{a}${\small{Department of Mechanical Engineering, Indian Institute of Technology Madras, Chennai - 600036, India}}
	\\ 
	$^{b}${\small{Institute for Applied Materials (IAM), Kalrsruhe Institute of Technology (KIT),}}\\ {\small{76344, Eggenstein-Leopoldshafen, Germany}}
	\\
	$^{c}${\small{School of Civil Engineering, The University of Sydney, NSW 2006, Australia}}\\ 
$^{\dagger}${\small{Corresponding author: ratna@iitm.ac.in}}}
\usepackage{amssymb}
\usepackage{amsmath}
\usepackage{txfonts}
\usepackage{graphicx}
\usepackage{subfigure}
\usepackage{color}
\usepackage[numbers,sort&compress,super]{natbib}
\usepackage{fancyhdr}
\usepackage{graphicx} 
\usepackage{booktabs} 
\usepackage{microtype} 
\usepackage{a4wide}
\usepackage{sectsty}
\usepackage{inputenc}
\usepackage{setspace}
\sectionfont{\MakeUppercase}
\newcommand{\addtext}[1]{{\color{black}{#1}}}

\makeatletter
\newcommand*{\citenumns}[2][]{%
  \begingroup
  \let\NAT@mbox=\mbox
  \let\@cite\NAT@citenum
  \let\NAT@space\NAT@spacechar
  \let\NAT@super@kern\relax
  \renewcommand\NAT@open{}%
  \renewcommand\NAT@close{}%
  \cite[#1]{#2}%
  \endgroup
}
\makeatother
\date{}
\begin{document}
\maketitle
\section*{Abstract}
Crushing strength of the breeder materials (lithium orthosilicate, $\rm{Li_4SiO_4}$ or OSi) in the form of pebbles to be used for EU solid breeder concept is investigated. The pebbles are fabricated using a melt-spray method and hence a size variation in the pebbles produced is expected. The knowledge of the mechanical integrity (crush strength) of the pebbles is important for a successful design of breeder blanket. In this paper, we present the experimental results of the crush (failure) loads for spherical OSi pebbles of different diameters ranging from $250~\mu$m to $800~\mu$m. The ultimate failure load for each size shows a Weibull distribution. Furthermore, the mean crush load increases with increase in pebble diameter. It is also observed that the level of opacity of the pebble influences the crush load significantly. The experimental data presented in this paper and the associated analysis could possibly help us to develop a framework for simulating a crushable polydisperse pebble assembly using discrete element method.

\section{Introduction}\label{sec:introduction}
Lithium Orthosilicate ($\rm{Li}_4\rm{SiO}_4$ or OSi) in the form of pebbles is the candidate breeder material for the \addtext{Helium Cooled Pebble Bed} (HCPB) breeder concept for European Union (EU) (Ref.~\citenumns{Boccaccini2009}). Thermo-mechanical integrity of the pebble bed is crucial for a safe and sustained fusion cycle.\citep{Ying2012} Experimental and numerical investigations to assess the breeder material properties~\citep{Reimann2000,Piazza2001a,Piazza2002,Buhler2002,An2005,Reimann2006,Knitter2007} and pebble bed packing structure~\citep{Reimann2003a,Reimann2008,Gan2010b} have been carried out in the past. It has also been shown that the pebble bed thermo-mechanical behaviour is strongly influenced by the individual pebble interactions and their packing structure in addition to the bulk material properties of the pebbles.\citep{Gan2010e,Zhao2012,Annabattula2012,Annabattula2012a,Zhao2013} The fabricated OSi pebbles have a size distribution and the size distribution has not been taken into consideration in the previous numerical studies except in~(Ref.~\citenumns{Annabattula2012}). The effect of pebble size distribution on the macroscopic behaviour has been studied using a polydisperse pebble assembly through Discrete Element Method (DEM)~(Ref.~\citenumns{Annabattula2012}). However, the study was focused on a non-crushable pebble assembly. But, under fusion relevant conditions, pebbles may fail possibly leading to termination of fusion fuel cycle. Recently, the studies on mechanics of a crushable pebble assembly have been reported albeit for a mono size pebble assembly.\citep{Gan2011,Zhao2012,Annabattula2012a,Zhao2013} For a more general understanding of the mechanics of pebble beds, the knowledge of the macroscopic response of a polydisperse crushable pebble assembly will be very useful in the design of pebble beds. However, such a study is not straight forward as in the case of mono size assemblies. We need a thorough understanding of crush (failure) load of individual pebbles as a function of pebble size which will be a necessary input for the numerical models mentioned in the foregoing discussion. Hence, in this paper we report the experimental details and the measured crush loads as a function of the pebble size for OSi pebbles.
\begin{figure*}
\begin{center}
\subfigure[]
{
\includegraphics[height=5cm]{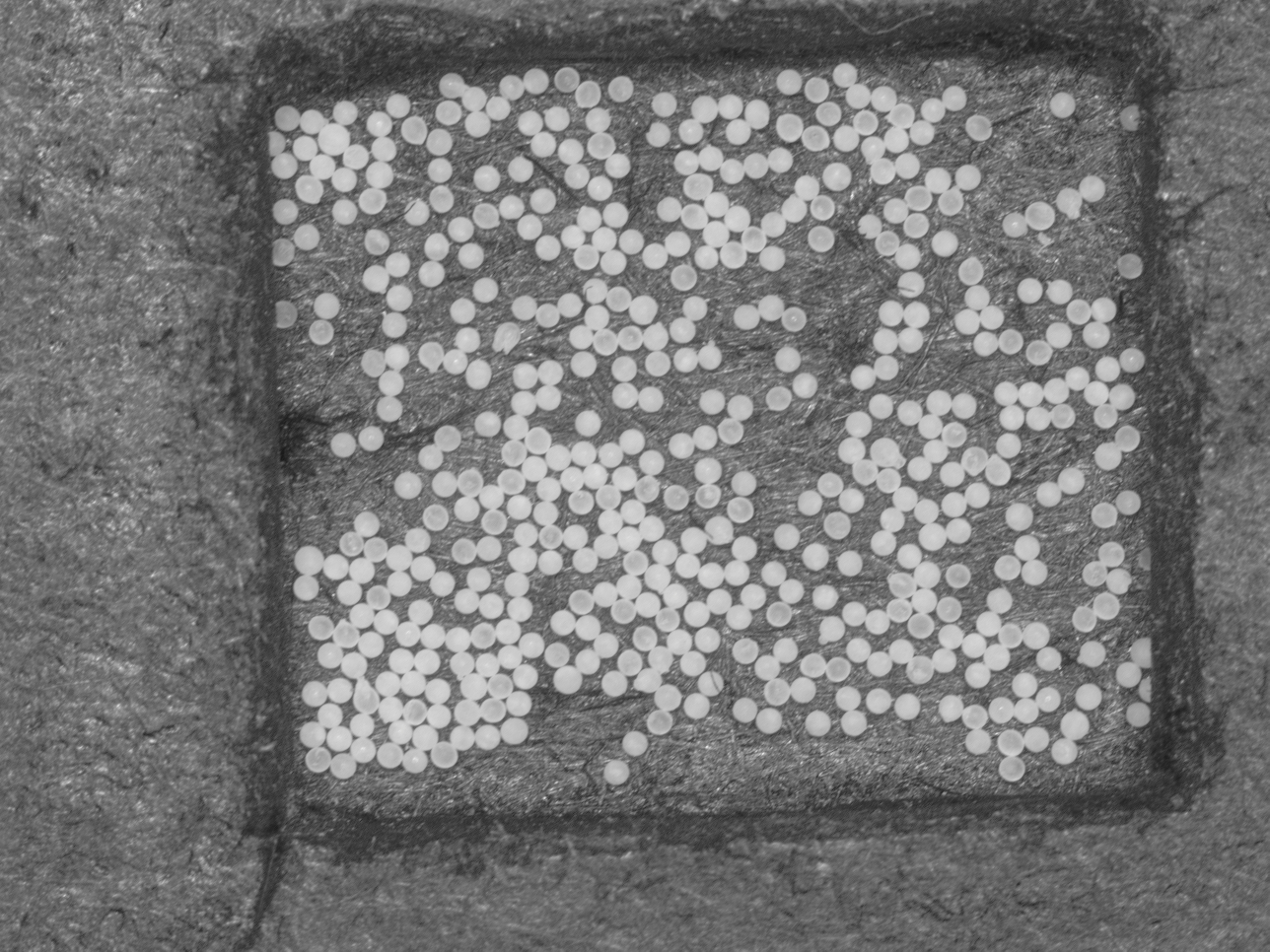}
\label{fig:250um-micrograph-optical}
}
\subfigure[]
{
\includegraphics[trim=100mm 60mm 80mm 60mm, clip, height=5cm]{FST13-737_Fig1.jpg}
\put(-25,25){{\color{white}{\bf{O}}}}
\put(-20,97){{\color{white}{\bf{T}}}}
\label{fig:250um-micrograph-transparency}
}
\end{center}
\caption{(a) Optical images of a batch of OSi pebbles of size 250 $\mu$m used for crush tests. (b) Zoomed view of the batch showing different transparency levels for pebbles. The transparent pebbles (marked ``T") fail at much lower load compared to opaque (or less transparent marked ``O") pebbles. Also, some of the pebbles are not completely spherical and hence the sphericity is also measured during the tests.}
\label{fig:250um-micrograph}
\end{figure*}
\section{Experimental Procedure}\label{sec:method}
Crush experiments have been conducted on OSi pebbles of diameter in the range of $250~\mu$m - $800~\mu$m at Fusion Materials Laboratory of Karlsruhe Institute of Technology (KIT), Germany. First, the pebbles are sieved into 10 different size groups with mean diameters of 250, 315, 355, 400, 450, 500, 560, 630, 710 and 800 (all in micrometer).  Then, the as received OSi pebbles are heated up to 300$^o$C for one hour in an inert gas (Nitrogen) environment to remove any moisture present. Individual pebbles are compressed \addtext{quasistatically} using a table top uniaxial testing machine between two compression platens made of BK7 glass. BK7 glass is chosen as the material for compression platens to reduce the effect of plastic deformation of platens on the failure load.\citep{Zhao2013} \addtext{The crush experiments have been conducted in a glove box at room temperature and the compression platens contact the pebble at the top and bottom in all the experiments.} The pebble size is measured before the crush test as the distance between the compression platens. The sphericity\footnote{\addtext{Sphericity in this study is defined as the ratio of minimum diameter to maximum diameter of a pebble measured in two orthogonal directions. The average of 60 measurements for each size is reported as the sphericity for the group shown in Table~\ref{table:sphericity}}} of the pebbles is measured through optical means by placing a layer of pebbles on a flat surface as shown in Fig.~\ref{fig:250um-micrograph-optical}. The average sphericity of the pebbles is approximately 0.95 (see Table~\ref{table:sphericity}) and hence the pebble size measured as the distance between compression plates is a reasonable estimate. The spread of the pebble size for a particular mean size is very small (less than 5\%) and hence in the results discussed in the following sections, only the mean size without error bars is considered. For each pebble size, 45 measurements have been made to take into account of the stochastic nature of the crush loads similar to the previous observations.\citep{Annabattula2012a} Furthermore, in Fig.~\ref{fig:250um-micrograph-transparency} two types of pebble surface morphology may be observed. The pebbles which are more transparent are marked as ``T" while relatively opaque pebbles are marked ``O". During the crush experiments, it has been observed that the transparent pebbles (marked ``T") show a very low crush load while the opaque (marked ``O") pebbles show an average crush load with some distribution. Hence, in the data analysis, the crush loads of transparent pebbles (show negligibly small crush load) is discarded. 
\begin{table*}
	\caption{Average Form Factor (sphericity) as a function of pebble size measured for 60 samples each size.}
	\label{table:sphericity}
	\begin{center}
	\begin{tabular}{lcccccccccc}\toprule
	Mean Pebble Diameter ($\mu$m) & 250 & 315 & 355 & 400 &450 &500 &560 &630 & 710 & 800\\ 
	\hline
	Form Factor (Sphericity) & 0.92 & 0.95 & 0.94 & 0.95 & 0.95 & 0.95 & 0.95 & 0.95 & 0.96 & 0.95\\
	\toprule
	\end{tabular}
\end{center}
\end{table*}
\section{Results and Discussion}\label{sec:results}
\begin{figure*}
	\begin{center}
		\subfigure[]
		{
			\includegraphics[width=7.0cm]{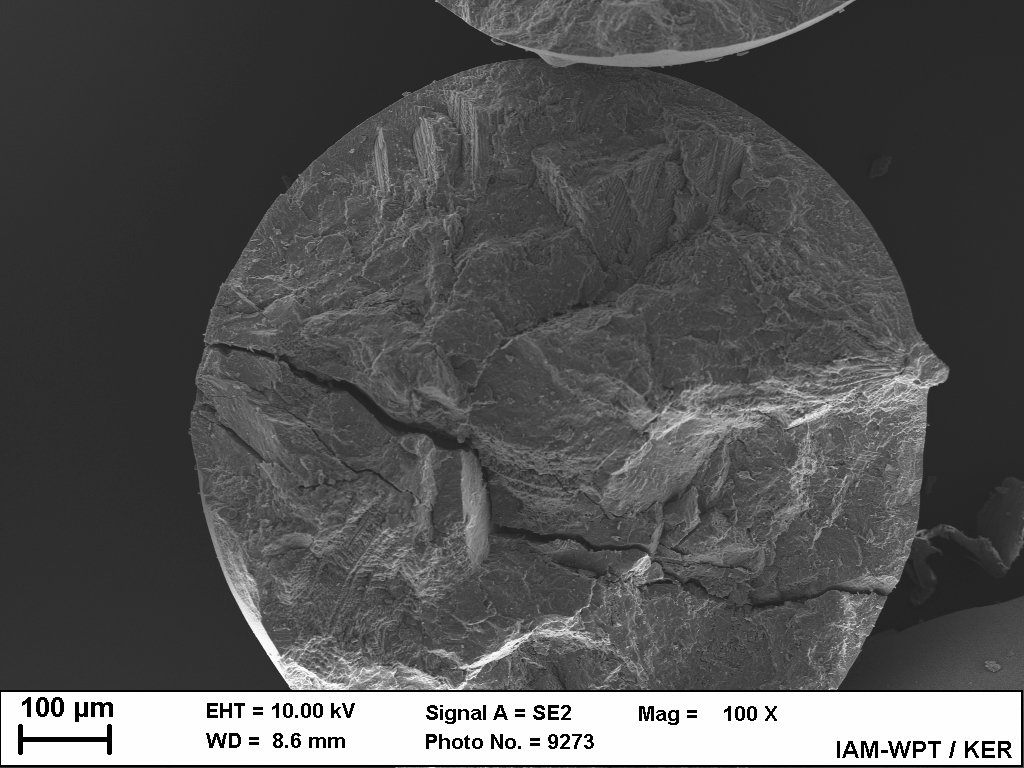}
			\label{fig:sem-image-failed-pebble}
		}
		\subfigure[]
		{
			\includegraphics[width=7.0cm]{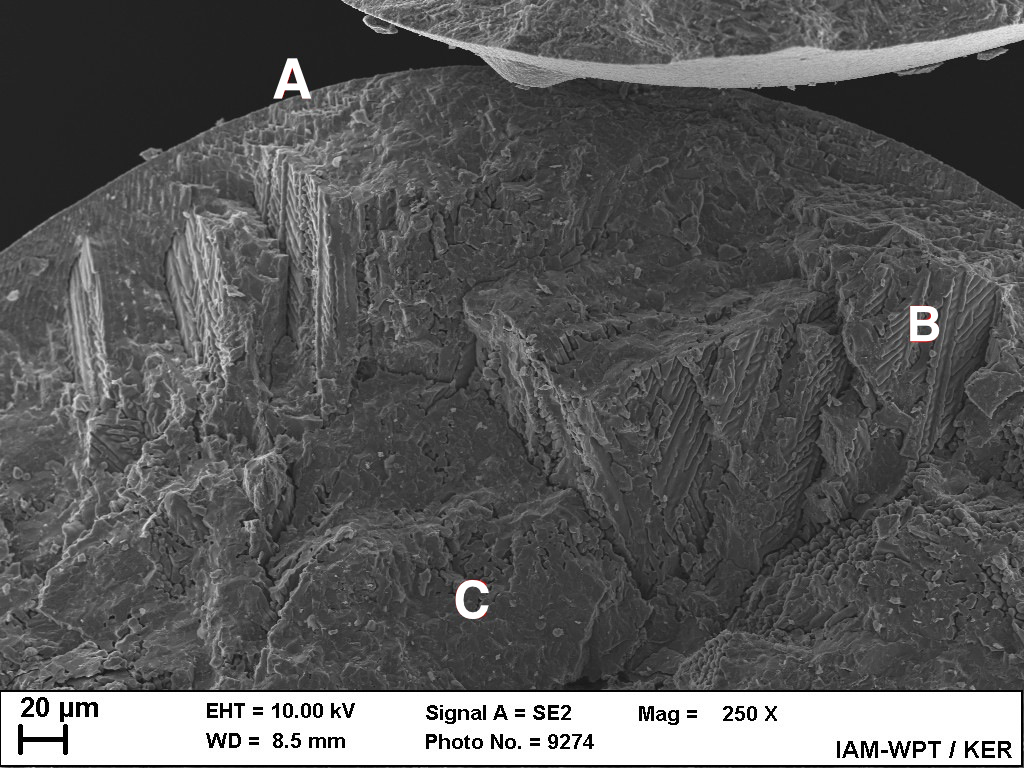}
			\label{fig:sem-image-failed-pebble-zoomed}
		}
		\subfigure[]
		{
			\includegraphics[width=7.0cm]{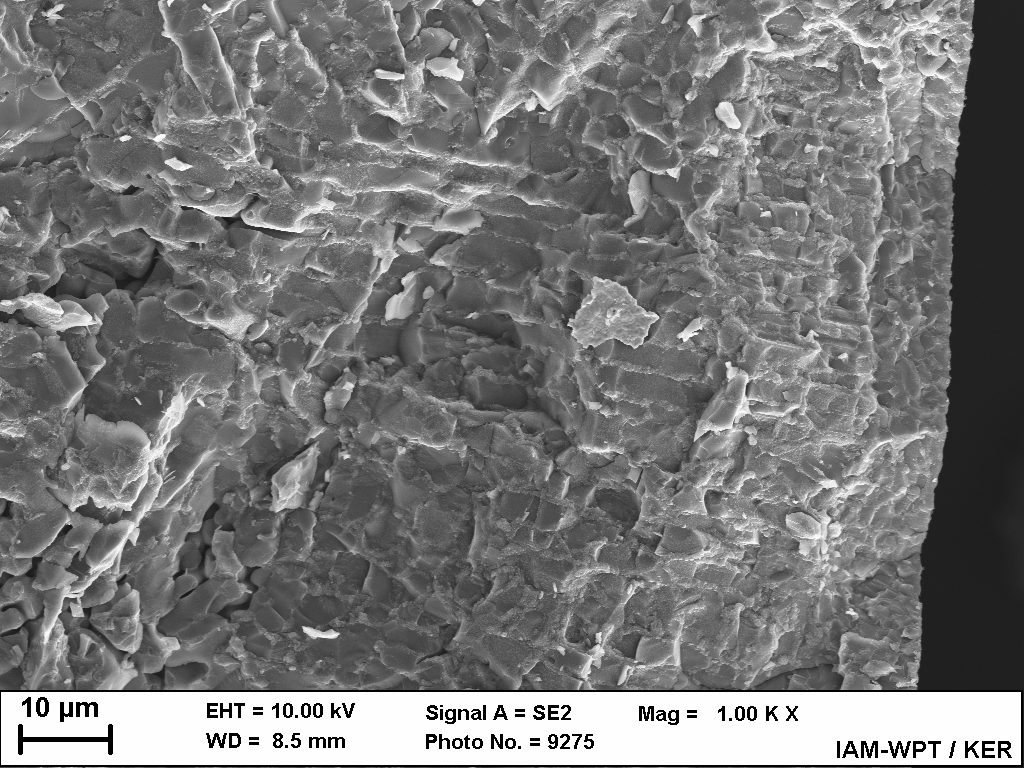}
			\put(-100,75){\color{white}{\Huge{A}}}
			\label{fig:sem-image-failed-pebble-RegionA}
		}
		\subfigure[]
		{
			\includegraphics[width=7.0cm]{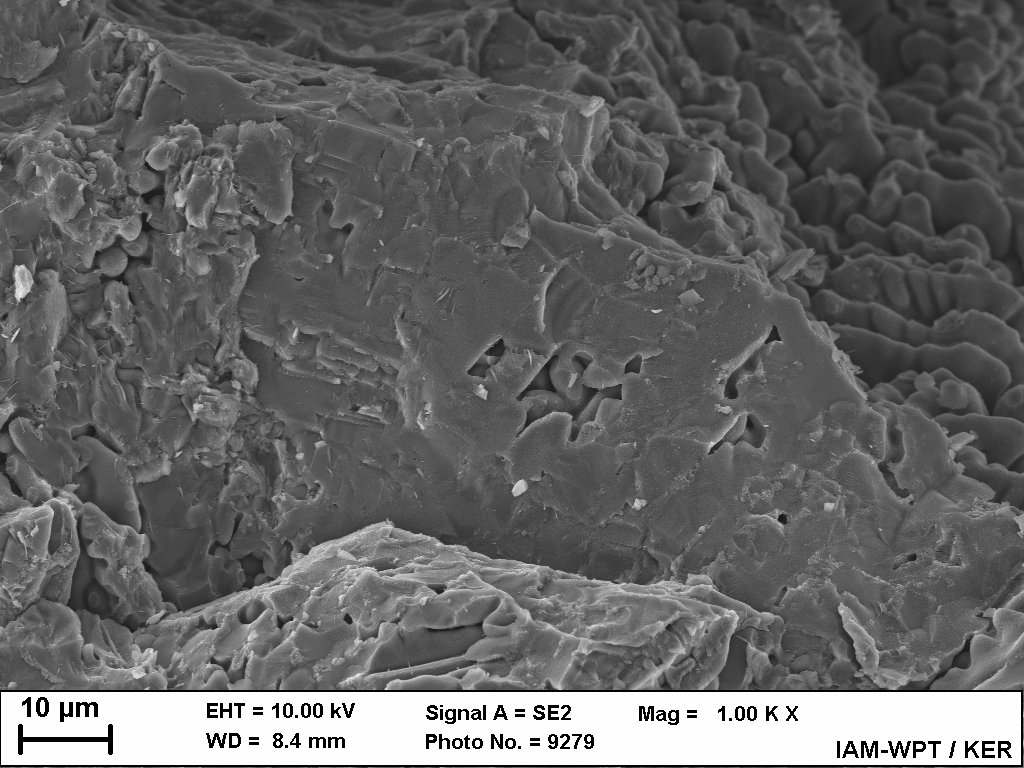}
			\put(-120,75){\color{white}{\Huge{C}}}
			\label{fig:sem-image-failed-pebble-RegionC}
		}
	\end{center}
	\caption{(a) SEM image of a typical failure surface of a OSi pebble. (b) Zoomed version of failure surface around top middle portion of the pebble shown in (a) describing various failure modes identified by region ``A" (slow crack growth), region ``B" (with voids) and region ``C" (clean fracture surface). \addtext{(c) Surface details of region ``A" and (d) region ``C".} }
\end{figure*}

In this section, we present the results obtained from the single pebble crush experiments followed by discussion. Fig.~\ref{fig:sem-image-failed-pebble} shows the SEM image of a typical OSi pebble after failure. The details of the failure surface are delineated in Fig.~\ref{fig:sem-image-failed-pebble-zoomed},\addtext{ Fig.~\ref{fig:sem-image-failed-pebble-RegionA} and Fig.~\ref{fig:sem-image-failed-pebble-RegionC}}. Here, three typical regions A, B and C are highlighted each depicting a different failure mechanism.\\
Region {\it{A}}: In this region the cracking of the pebble seems to be advancing quite slowly and definitely not in a catastrophic manner as we can observe a stepped surface \addtext{(see Fig.~\ref{fig:sem-image-failed-pebble-RegionA})} rather than a clean fracture surface typical of catastrophic failure. Although, such surfaces suggest some plastic deformation, it is not observed predominantly and hence plastic deformation may be ignored in these systems.\\
Region {\it{B}}: This type of micro structure is typical for OSi pebbles produced by a melt-spray process such as in this study.\citep{Knitter2007a,Kolb2011} Here the solidification of the melt leads to a flat void or pore within the pebble, compensating for the increase in density by crystallization. The void helps the crack to propagate resulting in a lower mechanical strength of the pebble.\\
Region {\it{C}}: This region shows the situation of a catastrophic failure of the pebble, i.e. the critical crack length is reached. We can observe a very clean fracture surface \addtext{(see Fig.~\ref{fig:sem-image-failed-pebble-RegionC})} without any steps as in region {\it{A}}. The crack propagates through the rest of the pebble at very high speed and the pebble breaks into few fragments.\\
From the above discussion, it can be clearly seen that the OSi pebble failure is brittle in nature. Furthermore, the crush load data follows a Weibull distribution (see~\citenumns{Annabattula2012a,Zhao2013}) typical of brittle failure in materials. \addtext{Details of the Weibull distribution for the experimental data is not presented here due to space limitations.} 
\par
\begin{figure*}
\begin{center}
\subfigure[]
{
\includegraphics[scale=1.0]{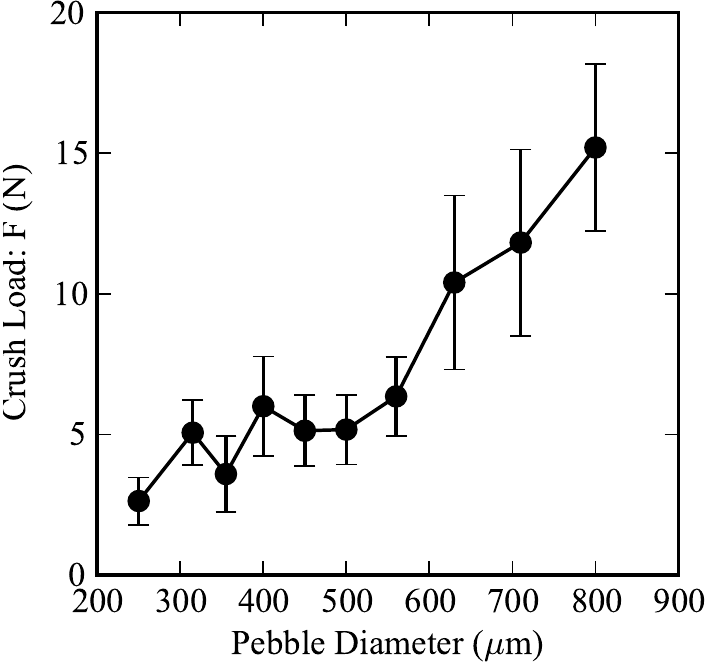}
\label{fig:crushload-vs-pebble-size}
}
\subfigure[]
{
\includegraphics[scale=1.0]{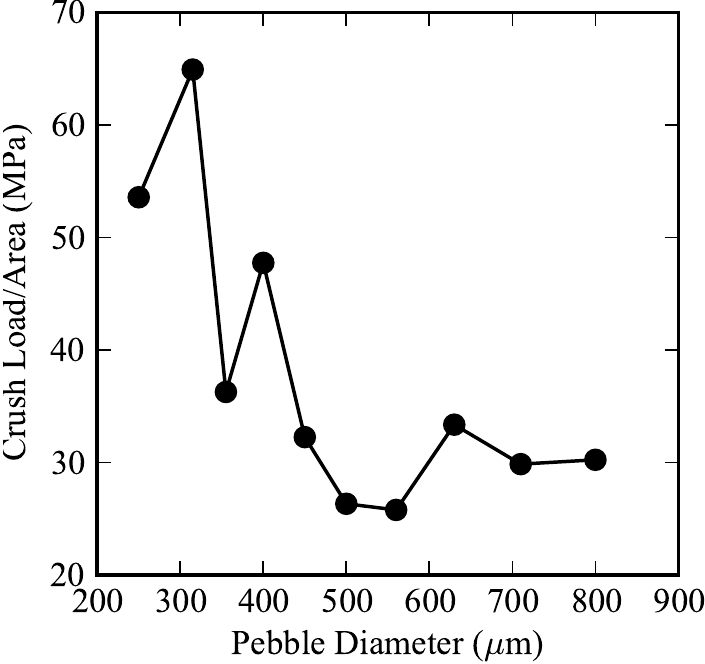}
\label{fig:crushstress-vs-pebble-size}
}
\end{center}
\caption{(a) Crush Load and (b) Equivalent Crush Stress for OSi pebbles as a function of pebble size.}
\end{figure*}

\begin{figure*}
\begin{center}
\includegraphics[width=8.0cm]{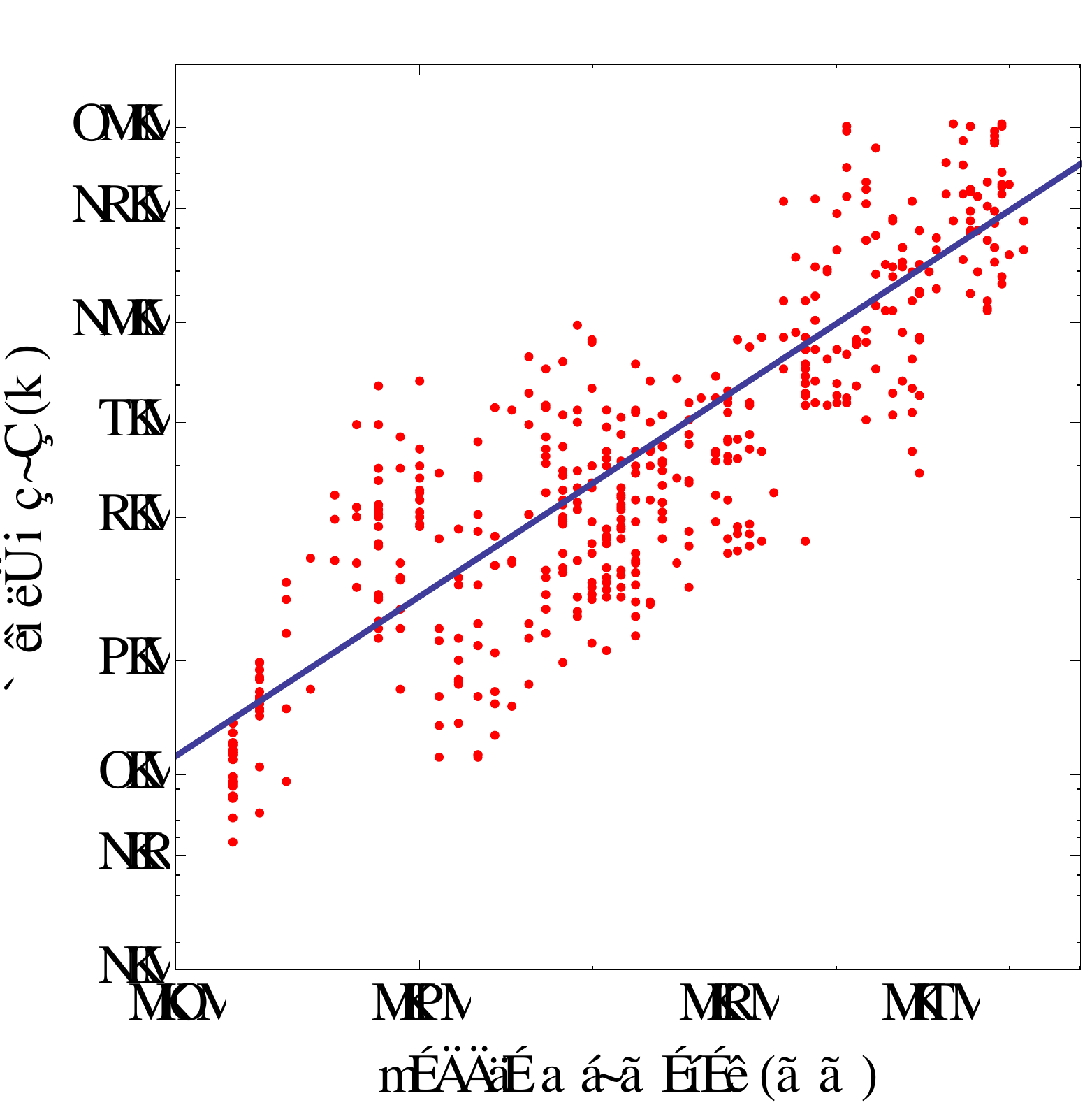}
\caption{Crush load as a function of pebble size (filled circles) and an approximate fit to the data for a failure energy $W_c^*=0.02~\rm{kJ/m}^2$.}
\label{fig:pebblecrush-fit}
\end{center}
\end{figure*}

 Fig.~\ref{fig:crushload-vs-pebble-size} shows the crush (failure) load\footnote{Crush load in the present analysis is defined as the point at which load suddenly drops in the load-displacement curve indicating the sudden fracture of pebbles.}($F$) as a function of pebble size.  Each data point represents an average of 45 measurements with a standard deviation as shown in the figure. The crush load increases with increase in pebble size. Also note that the standard deviation also increases with increase in pebble size. This increase in standard deviation may be related to increase in number of defects  and the size of defects with increase in pebble size. However, the data point corresponding to 355 $\mu$m shows a lower crush load compared to the expected trend. This is due to the presence of large fraction of transparent pebbles (indicated by ``T" in Fig.~\ref{fig:250um-micrograph-transparency}) for which the crush load is not very small unlike the cases in which such results are discarded as mentioned before. Fig.~\ref{fig:crushstress-vs-pebble-size} shows the variation of a measure of crush stress\footnote{The crush stress here should be looked as crush load per unit surface area and not as stress in physical sense} as a function of pebble size. Here, the crush stress is calculated by dividing the crush load with the area of a circle of same diameter as the pebble size. A size dependent crush stress that increases with decrease in pebble size can be observed.  The data point corresponding to 355 $\mu$m is an exception due to the presence of large number of transparent pebbles in the measurements as mentioned before. For characterizing the pebble failure a better measure is the critical failure energy, i.e. the stored elastic energy at the onset of failure of pebble during compression.\citep{Annabattula2012a,Zhao2012,Zhao2013} 
Critical failure energy ($W_c$) of a pebble in elastic contact is given by~(Ref.~\citenumns{Zhao2013})
\begin{equation}
	W_c = \sum_{i=1}^{N_c}\frac{1}{5}\left(\frac{9F_i^5}{16R_i^*E^{*^2}}\right)^\frac{1}{3}
	\label{eq:toughness}
\end{equation}
where $N_c$ is the number of contacts, $F_i$ is the normal contact force due to $i^{\rm{th}}$ contact. \addtext{The effective elastic modulus ($E^*$) and effective radius ($R^*$) are given by
\begin{eqnarray}
	E^* = \frac{1}{2}\left[\frac{E_p}{1-\nu_p^2}+\frac{E_{bk7}}{1-\nu_{bk7}^2}\right];\,\frac{1}{R^*} =\frac{1}{R_{p}}+\frac{1}{R_{bk7}}, 
\end{eqnarray}
where, $(E_p, E_{bk7}) = (90, 82)$~GPa and $(\nu_p, \nu_{bk7}) = (0.25, 0.206)$ are the Young's modulus and Poisson's ratio of the OSi pebble and BK7 glass materials, respectively. For the present case of a flat BK7 glass plate contacting OSi pebble, $R_{bk7}=\infty$ and hence $R^*=R_p$.} We also define another parameter critical failure energy per unit area ($W_c^*$) given by
\begin{equation}
	W_c^* = \frac{W_c}{\pi R^2}
	\label{eq:wcstar}
\end{equation}
The crush load experiments reported in this paper can be treated as mono-size pebbles with only a single contact pair. However, one needs to know the critical failure energy in order to predict a crush load versus pebble size variation as in Eq.~\ref{eq:toughness}. We hypothesise that the critical failure energy per unit area ($W_c^*$) of OSi pebbles doesn't depend on the size of the pebble and hence we can assume a constant mean failure energy per unit area to predict the crush load versus pebble size variation as in Eq.~\ref{eq:toughness}. We have already shown that the pebble failure is brittle (catastrophic) failure and there is no plastic deformation observed in these systems. Hence, the pebble fails when the strain energy release rate near the crack tip reaches the toughness (or critical failure energy per unit area) of the pebble material. Hence, we assume $W_c^* = 0.02~\rm{kJ/m}^2$ for the present analysis\footnote{This is the mean critical failure energy per unit area of the pebble with a diameter of $500~\mu$m reported in previous studies by~(Ref.~\citenumns{Annabattula2012a})}. Fig.~\ref{fig:pebblecrush-fit} shows the crush load of all the pebbles (all sizes) plotted against the pebble size in filled circles. The line in the figure is the fitted curve for the data with $W_c^* = 0.02~\rm{kJ/m}^2$ showing good fit with the experimental data. Note that the figure is plotted on log-log scale and hence the linear relation between crush load and pebble size should be understood in the right scale. Also, note that the critical failure energy reported for 500 $\mu$m pebbles is a not a single value but follows a Weibull distribution~\citep{Annabattula2012a} which is not taken into consideration in this analysis. \addtext{The transition from lab scale analysis of experiments presented in this paper to real DEM simulations involving pebbles of different sizes in contact needs further work on which the authors are currently working.}
\section{Conclusions}\label{sec:conclusions}
The crush load data for pebbles of 10 different sizes has been reported. The crush data for each size shows a Weibull distribution similar to previous studies. The crush load increases with increase in pebble size. The standard deviation of crush load also increases with increase in pebble size. This may be attributed to increase in number of defects and size of defects with increase in pebble size. The pebble crush data for different pebble sizes can be fitted with a single curve with an assumption of critical failure energy per unit area. The critical failure energy per unit area of $0.02~\rm{kJ/m}^2$ obtained for 500 $\mu$m pebble from the previous studies shows a reasonable agreement with the data. Hence, such a trend may be incorporated in future DEM simulations for a crushable polydisperse pebble assembly which is expected to give new insights to pebble bed thermo-mechanics.

\bibliographystyle{ans}
\bibliography{references}

\begin{thebibliography}{10}
\newcommand{\enquote}[1]{``#1''}

\bibitem{Boccaccini2009}
\MakeUppercase{L.~Boccaccini}, \MakeUppercase{J.-F. Salavy},
  \MakeUppercase{O.~Bede}, \MakeUppercase{H.~Neuberger},
  \MakeUppercase{I.~Ricapito}, \MakeUppercase{P.~Sardain},
  \MakeUppercase{L.~Sedano}, and \MakeUppercase{K.~Splichal}, \enquote{{The EU
  TBM systems: Design and development programme},} \emph{Fusion Engineering and
  Design}, \textbf{84}, \emph{2-6}, 333--337 (Jun. 2009).

\bibitem{Ying2012}
\MakeUppercase{A.~Ying}, \MakeUppercase{J.~Reimann},
  \MakeUppercase{L.~Boccaccini}, \MakeUppercase{M.~Enoeda},
  \MakeUppercase{M.~Kamlah}, \MakeUppercase{R.~Knitter},
  \MakeUppercase{Y.~Gan}, \MakeUppercase{J.~G. van~der Laan},
  \MakeUppercase{L.~Magielsen}, \MakeUppercase{P.~D. Maio},
  \MakeUppercase{G.~DellOrco}, \MakeUppercase{R.~K. Annabattula},
  \MakeUppercase{J.~T. {Van Lew}}, \MakeUppercase{H.~Tanigawa}, and
  \MakeUppercase{S.~van Til}, \enquote{{Status of ceramic breeder pebble bed
  thermo-mechanics R\&D and impact on breeder material mechanical strength},}
  \emph{Fusion Engineering and Design}, \textbf{87}, \emph{7-8}, 1130--1137
  (Aug. 2012).

\bibitem{Reimann2000}
\MakeUppercase{J.~Reimann}, \MakeUppercase{E.~Arbogast},
  \MakeUppercase{M.~Behnke}, \MakeUppercase{S.~Mueller}, and
  \MakeUppercase{K.~Thomauske}, \enquote{{Thermomechanical behaviour of ceramic
  breeder and beryllium pebble beds},} \emph{Fusion Engineering and Design},
  \textbf{49-50}, \emph{1-4}, 643--649 (Nov. 2000).

\bibitem{Piazza2001a}
\MakeUppercase{G.~Piazza}, \MakeUppercase{J.~Reimann},
  \MakeUppercase{E.~G\"{u}nther}, \MakeUppercase{R.~Knitter},
  \MakeUppercase{N.~Roux}, and \MakeUppercase{J.~Lulewicz}, \enquote{{Behaviour
  of ceramic breeder materials in long time annealing experiments},}
  \emph{Fusion Engineering and Design}, \textbf{58-59}, 653--659 (Nov. 2001).

\bibitem{Piazza2002}
\MakeUppercase{G.~Piazza}, \MakeUppercase{J.~Reimann},
  \MakeUppercase{E.~G\"{u}nther}, \MakeUppercase{R.~Knitter},
  \MakeUppercase{N.~Roux}, and \MakeUppercase{J.~Lulewicz},
  \enquote{{Characterisation of ceramic breeder materials for the helium cooled
  pebble bed blanket},} \emph{Journal of Nuclear Materials}, \textbf{307-311},
  811--816 (Dec. 2002).

\bibitem{Buhler2002}
\MakeUppercase{L.~B\"{u}hler} and \MakeUppercase{J.~Reimann}, \enquote{{Thermal
  creep of granular breeder materials in fusion blankets},} \emph{Journal of
  Nuclear Materials}, \textbf{307-311}, 807--810 (Dec. 2002).

\bibitem{An2005}
\MakeUppercase{Z.~An}, \MakeUppercase{A.~Ying}, and \MakeUppercase{M.~Abdou},
  \enquote{Experimental \& Numerical study of ceramic breeder pebble bed
  thermal deformation behavior,} \emph{Fusion Science and Technology},
  \textbf{47}, \emph{4}, 1101 (2005).

\bibitem{Reimann2006}
\MakeUppercase{J.~Reimann}, \MakeUppercase{G.~Piazza}, and
  \MakeUppercase{H.~Harsch}, \enquote{{Thermal conductivity of compressed
  beryllium pebble beds},} \emph{Fusion Engineering and Design}, \textbf{81},
  \emph{1-7}, 449--454 (Feb. 2006).

\bibitem{Knitter2007}
\MakeUppercase{R.~Knitter}, \MakeUppercase{B.~Alm}, and
  \MakeUppercase{G.~Roth}, \enquote{{Crystallisation and microstructure of
  lithium orthosilicate pebbles},} \emph{Journal of Nuclear Materials},
  \textbf{367-370}, 1387--1392 (Aug. 2007).

\bibitem{Reimann2003a}
\MakeUppercase{J.~Reimann}, \enquote{{Influence of pebble bed dimensions and
  filling factor on mechanical pebble bed properties},} \emph{Fusion
  Engineering and Design}, \textbf{69}, \emph{1-4}, 241--244 (Sep. 2003).

\bibitem{Reimann2008}
\MakeUppercase{J.~Reimann}, \MakeUppercase{R.~Pieritz},
  \MakeUppercase{C.~Ferrero}, \MakeUppercase{M.~Dimichiel}, and
  \MakeUppercase{R.~Rolli}, \enquote{{X-ray tomography investigations on pebble
  bed structures},} \emph{Fusion Engineering and Design}, \textbf{83},
  \emph{7-9}, 1326--1330 (Dec. 2008).

\bibitem{Gan2010b}
\MakeUppercase{Y.~Gan}, \MakeUppercase{M.~Kamlah}, and
  \MakeUppercase{J.~Reimann}, \enquote{{Computer simulation of packing
  structure in pebble beds},} \emph{Fusion Engineering and Design},
  \textbf{85}, \emph{10-12}, 1782--1787 (Dec. 2010).

\bibitem{Gan2010e}
\MakeUppercase{Y.~Gan} and \MakeUppercase{M.~Kamlah}, \enquote{{Discrete
  element modelling of pebble beds: With application to uniaxial compression
  tests of ceramic breeder pebble beds},} \emph{Journal of the Mechanics and
  Physics of Solids}, \textbf{58}, \emph{2}, 129--144 (Feb. 2010).

\bibitem{Zhao2012}
\MakeUppercase{S.~Zhao}, \MakeUppercase{Y.~Gan}, and \MakeUppercase{M.~Kamlah},
  \enquote{{Spherical ceramic pebbles subjected to multiple non-concentrated
  surface loads},} \emph{International Journal of Solids and Structures},
  \textbf{49}, \emph{3-4}, 658--671 (Feb. 2012).

\bibitem{Annabattula2012}
\MakeUppercase{R.~K. Annabattula}, \MakeUppercase{Y.~Gan}, and
  \MakeUppercase{M.~Kamlah}, \enquote{{Mechanics of binary and polydisperse
  spherical pebble assembly},} \emph{Fusion Engineering and Design},
  \textbf{87}, \emph{5-6}, 853--858 (Aug. 2012).

\bibitem{Annabattula2012a}
\MakeUppercase{R.~K. Annabattula}, \MakeUppercase{Y.~Gan},
  \MakeUppercase{S.~Zhao}, and \MakeUppercase{M.~Kamlah}, \enquote{{Mechanics
  of a crushable pebble assembly using discrete element method},} \emph{Journal
  of Nuclear Materials}, \textbf{430}, \emph{1-3}, 90--95 (Nov. 2012).

\bibitem{Zhao2013}
\MakeUppercase{S.~Zhao}, \MakeUppercase{Y.~Gan}, \MakeUppercase{M.~Kamlah},
  \MakeUppercase{T.~Kennerknecht}, and \MakeUppercase{R.~Rolli},
  \enquote{{Influence of plate material on the contact strength of
  $\rm{Li}_4\rm{SiO}_4$ pebbles in crush tests and evaluation of the contact
  strength in pebble–pebble contact},} \emph{Engineering Fracture Mechanics},
  \textbf{100}, 28--37 (Mar. 2013).

\bibitem{Gan2011}
\MakeUppercase{Y.~Gan}, \MakeUppercase{M.~Kamlah},
  \MakeUppercase{H.~Riesch-Oppermann}, \MakeUppercase{R.~Rolli}, and
  \MakeUppercase{P.~Liu}, \enquote{{Crush probability analysis of ceramic
  breeder pebble beds under mechanical stresses},} \emph{Journal of Nuclear
  Materials}, \textbf{417}, \emph{1-3}, 706--709 (2011).

\bibitem{Knitter2007a}
\MakeUppercase{R.~Knitter} and \MakeUppercase{B.~Lobbecke},
  \enquote{{Reprocessing of lithium orthosilicate breeder material by
  remelting},} \emph{Journal of Nuclear Materials}, \textbf{361}, \emph{1},
  104--111 (Mar. 2007).

\bibitem{Kolb2011}
\MakeUppercase{M.~Kolb}, \MakeUppercase{R.~Knitter},
  \MakeUppercase{U.~Kaufmann}, and \MakeUppercase{D.~Mundt}, \enquote{Enhanced
  fabrication process for lithium orthosilicate pebbles as breeding material,}
  \emph{Fusion Engineering and Design}, \textbf{86}, \emph{9-11}, 2148 -- 2151
  (2011).

\end{thebibliography}
\end{document}